\documentclass[onecolumn,amsmath,amssymb]{qm2008_abs}
\usepackage{graphicx}
\usepackage{dcolumn}
\usepackage{bm}
\begin{document}

\title{{\Large Strongly Coupled Quark-Gluon Plasma:\\[1mm] 
Equation of State near {\boldmath$T_c$}\footnote{Contribution to
Quark Matter 2008, February 4-10, 2008, Jaipur, India.}}}

\bigskip
\bigskip
\author{\large M.~Bluhm$^{\,1,}$}
\email{m.bluhm@fzd.de}
\author{\large B.~K\"ampfer$^{\,1,2}$}
\affiliation{$^1$ Forschungszentrum Dresden-Rossendorf, 
PF 510119, 01314 Dresden, Germany \\
$^2$ Institut f\"ur Theoretische Physik, TU Dresden, 
01062 Dresden, Germany}
\bigskip
\bigskip

\begin{abstract}
\leftskip 1.0cm
\rightskip 1.0cm
We test the quark mass dependence implemented in the 
quasiparticle dispersion relations of our quasiparticle 
model for the QCD equation of state by comparing with 
recently available lattice QCD data near $T_c$ employing almost 
physical quark masses. In addition, we emphasize the 
capability of our model to successfully describe lattice 
QCD results for imaginary chemical potential and to 
analytically continue the latter to real chemical potential. 
\end{abstract}

\maketitle

\section{Introduction}

The equation of state of strongly interacting matter is of
paramount interest for understanding and describing the dynamics of
relativistic heavy-ion collisions and the early universe and
compact stellar objects as well. The grand canonical potential
$\Omega$ depends on a set of intensive quantities (temperature $T$,
various chemical potentials accumulated in $\vec \mu$), and
parameters like quark masses, 
flavor number $N_f$ and color number $N_c$. For QCD as theory of
hadrons and quarks and gluons these parameters are fixed, but it is
instructive to study their impact on $\Omega$, given the complexity
of QCD. Of particular interest is $\Omega$ near the demarcation line
(phase boundary) $T_c(\vec \mu)$ of hadrons vs. quarks and
gluons. This region is probed in heavy-ion collisions \cite{Bluhm:2007b} 
and is traversed in the cosmic evolution \cite{Henry_BK}.


To parameterize the equation of state near $T_c$ we employ a
quasiparticle model 
(QPM) \cite{Peshier:1994,Peshier:1996,Peshier:2000,
Peshier:2002,Bluhm:2004,Bluhm:2007b,Bluhm:2007a} 
which is adjusted to available
lattice QCD data. The striking simplicity of the model consists in the
expressions for net baryon density $n_B$ and entropy density 
$s=\sum_{i=q,g}s_i$ reading 
\begin{equation}
n_B(T, \mu_q) = \frac{N_cN_f}{3\pi^2}\int_0^\infty dk k^2 
\left(\frac{1}{e^{(\omega_q-\mu_q)/T}+1} - 
\frac{1}{e^{(\omega_q+\mu_q)/T}+1}\right) \,,
\label{eq.1}
\end{equation}
\begin{equation}
s_i(T, \mu_q) = \epsilon_i\frac{d_i}{\pi^2}\int_0^\infty dk k^2 
\left(\ln \left[1+\epsilon_ie^{-(\omega_i-\mu_i)/T}\right] + 
\epsilon_i\frac{(\omega_i-\mu_i)/T}{e^{(\omega_i-\mu_i)/T}+\epsilon_i} 
+ 
[\mu_i \rightarrow -\mu_i]
\right) \,,
\label{eq.2}
\end{equation}
where $d_q=N_cN_f$, $d_g=N_c^2-1$, $\epsilon_q=1$, $\epsilon_g=-1$ 
and $\mu_g\equiv 0$. 
The pressure $p = - \Omega /V$, where $V$ is the volume, 
is to be calculated consistently with 
Eqs.~(\ref{eq.1}, \ref{eq.2}), cf.\ \cite{Bluhm:2007b}.
We restrict here our attention onto 
considering one independent chemical potential $\mu_q$.
$\omega_{q,g}^2 = k^2 + \Pi_{q,g}(T, \mu_q)$ are the
dispersion relations for the quasiparticle excitations with
self-energies $\Pi_{q,g}(T, \mu_q)$ in line with 1-loop
approximations
and a procedure to include nonzero quark masses according to~\cite{Pisarski} 
\begin{equation}
\label{equ:self}
\Pi_{q,g} = m_{q,g}^2 + 2 m_{q,g} \hat{\omega}_{q,g} + 
2 \hat{\omega}_{q,g}^2 \,\,,
\end{equation}
where $m_g = 0$. 
In the quark sector,
the rest masses $m_q$ contain a ''true'' rest mass term
$m_q^{(0)}$ and an artificial ''lattice'' mass term 
$\xi_q T$ introduced for calculational purposes on the lattice,
$m_q^2 = m_q^{(0) \, 2} + \xi_q^2 T^2$. 
The interaction generates the self-energy contributions
$\hat{\omega}_{q,g}^2=G^2(\alpha_{q,g}T^2+\beta_{q,g}\mu_q^2)$ with 
$\alpha_q=\frac16$, $\beta_q=\frac{1}{6\pi^2}$, 
$\alpha_g=\frac{1}{24}(2N_c+N_f)$ and $\beta_g=\frac{N_f}{8\pi^2}$. 
$G^2$ is an effective coupling
strength parametrized at $\mu_q=0$ via 
\begin{equation}
G^2(T) = \left\{
    \begin{array}{l}
      \!\! G^2_{\rm 2-loop} (\zeta(T)), \quad T{\,\ge\,}T_c,
      \\[3mm]
      \!\! G^2_{\rm 2-loop}(\zeta(T_c)) + b \left(1{-}\frac{T}{T_c}\right), \ 
      T{\,<\,}T_c 
    \end{array}
  \right.
\end{equation}
with $\zeta(T)=\lambda(T-T_s)/T_c$, and 
approaches smoothly the perturbative region at large temperatures.
Near $T_c$, $G^2$ becomes large, 
and the shift parameter $T_s$ regulates the coupling.
Below $T_c$, the coupling changes drastically its behavior. 

The model may be considered as a resummed
expression for the thermodynamic potential, as a formal power
expansion in $G^2$ generates an infinite series of terms, including 
also a term proportional to 
the plasmon term, for instance. It goes beyond a
perturbative expansion scheme as result of the effective coupling $G^2$, which
also may repair possible shortcomings of the 1-loop inspired parameterizations 
of the dispersion relations.

In this form, 
the model does not contain critical point
(cf.\ \cite{Karsch:2002}) 
or color-flavor locking effects (cf.\ \cite{Rischke}).

Despite of its simplicity, the equation of state for real and 
imaginary chemical potential as well as various susceptibilities are 
described very well \cite{Bluhm:2007c,Bluhm:2008} in the QPM. 
Here, we describe one new aspect of our
model, namely a naive chiral extrapolation of the equation of state. We
emphasize also the capability to extrapolate lattice QCD results from 
imaginary $\mu_q$ to real $\mu_q$. We focus on
the region $T \sim T_c$, where interaction effects are strong.
It is the region
of the strongly coupled quark-gluon plasma 
\cite{Shuryak:2004,Gyulassy:2005,Teaney:2003} 
presently explored experimentally at RHIC and in near future at LHC 
and later on at FAIR.

\section{Chiral extrapolation \label{sec:2}}

The QPM parametrization of the lattice QCD results \cite{Karsch:2003} 
for $N_f=2+1$ with fairly large quark masses, i.e.
$\xi_{u,d} = 0.4$ and $\xi_s = 1$, 
was already presented in \cite{Bluhm:2007b} 
for the scaled pressure $p/T^4$ at $n_B=0$.
The QPM parameters read
$\lambda = 7.8$, 
$T_s/T_c = 0.8$, 
$b = 347$ 
for $m_{u,d,s}^{(0)} = 0$. 

\begin{figure}[ht]
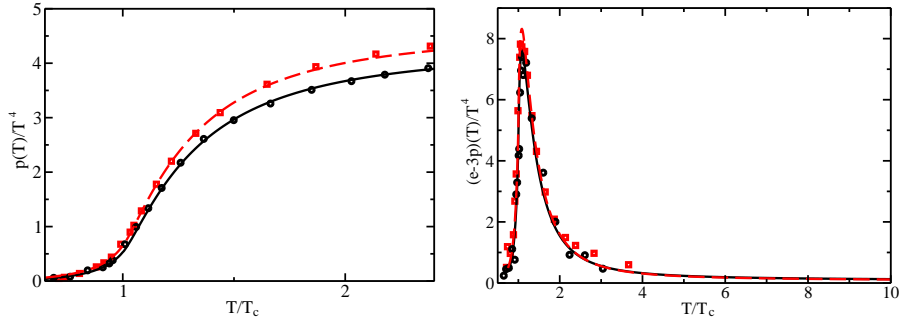

\centering
\vskip 4mm
\includegraphics[width=.38\textwidth,angle=0]{pres46_5.eps}
\hskip 2mm
\includegraphics[width=.39\textwidth,angle=0]{intmessure2.eps}
\parbox[h]{.8\linewidth}{
\caption{\label{fig:pNf2+1extrap}
Exploration of the quark mass dependence in the quasiparticle 
dispersion relations. Left: 
Circles and squares  
exhibit lattice QCD results for the scaled pressure $p(T)/T^4$ 
for $N_f=2+1$ at $n_B=0$ for $m_{u,d} = 0.4\,T$, $m_s=T$ \cite{Karsch:2003}
and almost physical quark masses \cite{Cheng:2007}, respectively. 
The black solid curve 
shows the QPM parametrization \cite{Bluhm:2007b} of lattice QCD data \cite{Karsch:2003}. 
The red dashed curve exhibits the QPM result when changing the quark 
masses to $m_{u,d} = 0.015\,T$, $m_s=10\,m_{u,d}$, 
corresponding to \cite{Cheng:2007}, leaving the parameters 
in $G^2$ unchanged but rendering $B(T_c)/T_c^4$ from 
$0.54$ to $0.76$. 
Right: Comparison of QPM with lattice QCD 
results for the scaled interaction 
measure $(e-3p)/T^4$ (line codes and symbols as in the 
left panel).}
}
\end{figure}

Now we try to accomplish a chiral extrapolation 
by means of Eq.~(\ref{equ:self}). 
Neglecting naively a conceivable 
dependence of $\lambda$, $T_s/T_c$ and $b$ on $m_q$, 
the extrapolated results are exhibited in the left panel 
of Figure~\ref{fig:pNf2+1extrap} (dashed curve) 
for $\xi_{u,d} = 0.015$ and $\xi_s = 0.15$, corresponding to
the set-up in \cite{Cheng:2007}.
These lattice QCD results \cite{Cheng:2007} 
(squares) are astonishingly well reproduced, however, by the
price of changing the pressure integration constant, $B(T_c)$, 
which needs to be 
readjusted in order to generate the small pressure below $T_c$. 
We note that putting $\xi_{u,d,s} = 0$ in the strikt chiral limit
does not change noticeably the dashed curve on the scale exhibited in
Figure 1.

Another important test \cite{Hung:1995} of the suitability 
of our hydrodynamic EoS is the interaction measure $(e-3p)$, 
where $e$ denotes the energy density. As shown in the right 
panel of Figure~\ref{fig:pNf2+1extrap}, the QPM for almost physical 
quark masses (i.e., $\xi_{u,d} = 0.015$ and $\xi_s = 0.15$)
faithfully reproduces corresponding lattice QCD 
data \cite{Cheng:2007} of $(e-3p)/T^4$. The peak, which is 
related to the softest point in the QCD equation of state, 
is located at $T/T_c=1.08$. For larger temperatures, the 
interaction measure approaches logarithmically zero 
according to the temperature dependence in the effective 
coupling $G^2$, though, is close to the conformal limit 
$e=3p$ already for $T/T_c\ge 10$. 

\section{Imaginary chemical potential \label{sec:3}}

\begin{figure}[ht]
\vskip 1mm
\centering
\vskip 1mm
\includegraphics[width=.39\textwidth,angle=0]{dens6_5.eps}
\hskip 1mm
\includegraphics[width=.385\textwidth,angle=0]{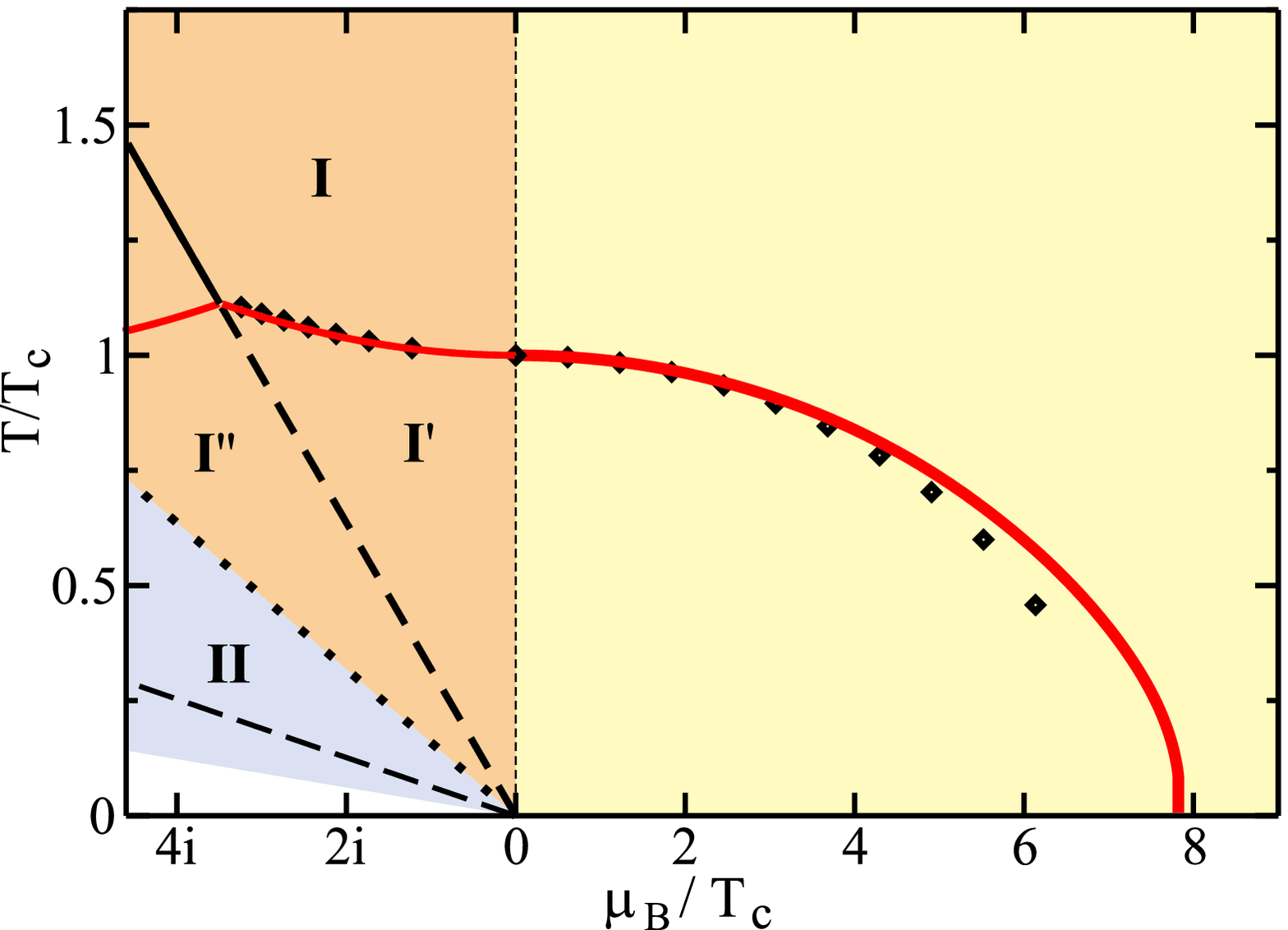}\\[1mm]
\parbox[h]{.8\linewidth}{
\caption[Scaled net quark number density for imaginary 
chemical potential]
{\label{fig:densnqimmu}
Left: Comparison of QPM~\cite{Bluhm:2007c} (solid curves) 
with continuum estimated lattice QCD 
results~\cite{D'Elia:2004,D'Elia:2007} (symbols) for the imaginary part 
of the scaled net quark number density $n_q/T^3$ for $N_f=4$ as a 
function of $\mu_i/T$ 
for temperatures $T=1.1, 1.5, 2.5, 3.5\,T_c$ (diamonds, 
circles, squares and triangles, respectively). The 
discontinuity in $n_q/T^3$ takes place 
at the Roberge-Weiss critical chemical potential 
$\mu_c/T=\pi/3$. For $T\ge 1.5\,T_c$, $n_q/T^3$ as a 
function of $\mu_i/T$ exhibits an interesting scaling 
behavior with $T$. 
Right: QCD phase diagram for $N_f=4$ at imaginary and real 
$\mu_B$. Diamonds represent a polynomial 
fit to the lattice QCD data in \cite{D'Elia:2003} in the sector of 
imaginary chemical potential and its analytic continuation, while the
solid curve is the QPM result for the parametrization 
of the pseudo-critical line
from \cite{Bluhm:2007c}. See text for details
of the phase structure in the imaginary sector.
}}
\end{figure}

In this section, QCD thermodynamics for $N_f=4$ at nonzero 
imaginary chemical potential $\mu_q \equiv i\mu_i$ is considered.
The net quark number density $n_q=3\,n_B$ reads in the QPM \cite{Bluhm:2007c}
\begin{equation}
\label{equ:densquarkimmu2}
    n_q(T,i\mu_i) = i\frac{d_q}{\pi^2}\int_0^\infty dk k^2 
    \left(\frac{e^{\omega_q/T}\sin (\mu_i/T)}
    {e^{2\omega_q/T}+2e^{\omega_q/T}\cos (\mu_i/T)+1}\right) \,, 
\end{equation}
which is purely imaginary and an odd function in $\mu_i$. 
In Figure~\ref{fig:densnqimmu} (left panel), QPM results for 
$-i n_q/T^3$, employing the parametrization from 
\cite{Bluhm:2007c}, are compared with 
the lattice QCD results \cite{D'Elia:2004,D'Elia:2007}. 
In particular, the pronounced bending for $T=1.1\,T_c$ close 
to $\mu_c/T=\pi/3$, which signals the onset of the 
first-order Roberge-Weiss transition \cite{Roberge:1986}, 
is accurately reproduced representing a sensible test of 
the QPM at nonzero $n_B$. In the QPM, the bending is driven 
by the increasing $\mu_i$ dependence in the quasiparticle dispersion 
relation $\omega_q$ close to $\mu_c$. 
For temperatures $T\ge 1.5\,T_c$ we observe 
an independence of $n_q/T^3$ considered as a function of 
$\mu_i/T$ on the explicit value of $T$. This independence 
follows from Eq.~(\ref{equ:densquarkimmu2}) as long as $\omega_q$ 
is approximately independent of $\mu_i$. 

The phase boundary $T_c(\mu_B)$ for $N_f=4$ is shown in the 
right panel of Figure~\ref{fig:densnqimmu} for imaginary 
as well as for real baryo-chemical potential $\mu_B=3\mu_q$, which is 
accessible by analytic continuation. Within the QPM, 
an estimate for $T_c(\mu_B)$ follows as self-consistent 
solution of a partial differential equation based on Maxwell's 
relation emerging at $T_c(\mu_B=0)$ (cf.~\cite{Bluhm:2007c}). 
$T_c(\mu_B)$ for imaginary $\mu_B$ and the 
first Roberge-Weiss transition line cross each other at 
$T^*/T=1.112$ and $(\mu_B^*)^2/T_c^2=-12.214$ being close 
to the lattice QCD results 
$T^*/T=1.095$ and $(\mu_B^*)^2/T_c^2=-11.834$ 
\cite{D'Elia:2003,D'Elia:2004}. 

The features of the phase diagram in the imaginary chemical potential sector
can be described as follows.
Dashed curves represent the analytic 
sections of the first two Roberge-Weiss transition lines 
$\mu_B^2/T_c^2=-T^2\pi^2(2k+1)^2/T_c^2$ (here $k=1,2$) turning 
into first-order transitions (solid section) while the 
dotted curve shows the first $\mathcal{Z}_3$ center 
symmetry line $\mu_B^2/T_c^2=-4T^2\pi^2/T_c^2$. 
The Roberge-Weiss periodicity \cite{Roberge:1986} implies 
that sectors between $\mu_i/T=2\pi k/3$ and $2\pi(k+1)/3$ 
(sector II for $k=1$) are repeated copies of sector I 
between $\mu_i/T=0$ and $2\pi/3$. The subsector 
between $\mu_i/T=\pi/3$ and $2\pi/3$ (sector I'') is a 
reflected copy of the subsector between $\mu_i/T=0$ and 
$\pi/3$ (sector I') mirrored at the first Roberge-Weiss 
transition line.

\section{Summary \label{sec:4}}

In summary we show that the quasiparticle model with the 
chosen dispersion relations for quarks and gluons accounts 
fairly well for the quark mass dependence in the QCD 
equation of state near $T_c$ as delivered by selected lattice QCD results.
We emphasize further that lattice QCD calculations at imaginary
chemical potential, avoiding the sign problem of the fermion determinant,
give valuable information which can be continued to real
chemical potential within our model.
The feasibility of exploring larger net baryon densities 
is relevant for future heavy-ion experiments at FAIR.\\[3mm]
The work is supported by 06DR136, GSI-FE, and EU I3HP.


\end{document}